# Vortex soliton tori with multiple nested phase singularities in dissipative media

Olga V. Borovkova[1], Valery E. Lobanov[1], Yaroslav V. Kartashov[1,2], and Lluis Torner[1]

[1]*ICFO-Institut de Ciencies Fotoniques, and Universitat Politecnica de Catalunya, Mediterranean Technology Park, 08860 Castelldefels (Barcelona), Spain*

[2]*Institute of Spectroscopy, Russian Academy of Sciences, Troitsk, Moscow Region, 142190, Russia*

We show the existence of stable two- and three-dimensional vortex solitons carrying multiple, spatially separated, single-charge topological dislocations nested around a vortex-ring core. Such new nonlinear states are supported by elliptical gain landscapes in focusing nonlinear media with two-photon absorption. The separation between the phase dislocations is dictated mostly by the geometry of gain landscape and it only slightly changes upon variation of the gain or absorption strength.

Optical vortices are light beams carrying a nonzero angular momentum and having a nontrivial phase distribution around a topological phase dislocation [1]. They are ubiquitous physical entities that find applications in diverse areas, including optical trapping, microscopy and quantum information, to name a few [2, 3]. Nonlinearity of the material offers the possibility to achieve propagation of vortex beams (vortex solitons) without distortion [4], but at the same time it often leads to self-destructive azimuthal instabilities of the host beams. Such instabilities may be eliminated in materials with competing nonlinearities [5,6], nonlocal materials [7], or in the presence of optical lattices [8-10]. Dissipative media [11,12] also may support stable vortex solitons due to the competition between various dissipative mechanisms, as it was shown for laser amplifiers [13], Ginzburg-Landau systems [14-16], and Bose-Einstein condensates [17,18].

Recently a new mechanism for stabilization of dissipative solitons, based on the spatial localization of gain, was put forward. In spatially localized gain landscapes the instability of the background surrounding the amplifying region is suppressed, leading to generation of stable one-dimensional solitons in Bragg gratings and waveguides [19,20], optical lattices [21,22], and Bose-Einstein condensates with complex potentials [23]. Moreover, in the ring-shaped gain landscapes imprinted in the focusing Kerr medium with two-photon absorption also the azimuthal modulation instabilities can be suppressed and the formation of stable stationary [24] and rotating [25] multicharged vortex solitons becomes possible.

While in conservative uniform media vortex solitons with any topological charge have radially symmetric shapes and carry a single phase dislocation in the center, in media where the refractive index or the nonlinearity are transversally inhomogeneous or depend on the sample geometry, it becomes possible to form nonconventional vortex soliton states featuring several, spatially separated phase dislocations nested in a common vortex core. Such phenomenon has been studied in conservative thermal materials [26] and in optical lattices [27]. In this paper we show that elliptical two-dimensional (2D) and also three-dimensional (3D) gain landscapes with strong nonlinear absorption added to cubic nonlinear media also support nonconventional stable vortex solitons and vortex tori that carry *several spatially separated single-charge dislocations*. Such vortex states are shown to be stable and robust, and the distance between the individual phase dislocations is dictated mostly by the ellipticity of the gain landscape.

We consider the propagation of a light beam along the $\xi$-axis in a cubic nonlinear medium with two-photon absorption and nonuniform gain that can be described by the nonlinear Schrödinger equation for the dimensionless light field amplitude $q$. For the sake of computational simplicity, first we analyze 2D geometries, where the governing equation writes

$$i\frac{\partial q}{\partial \xi} = -\frac{1}{2}\left(\frac{\partial^2 q}{\partial \eta^2} + \frac{\partial^2 q}{\partial \zeta^2}\right) - q|q|^2 + i\gamma(\eta,\zeta)q - i\alpha q|q|^2, \tag{1}$$

Here $\xi$ is the propagation distance normalized to the diffraction length; $\eta$ and $\zeta$ are the transverse coordinates normalized to the characteristic transverse scale; $\alpha$ is the strength of two-photon absorption; the gain shape is described by the function $\gamma = p_i \exp[-(r-r_c)^2/a^2]$, where $p_i$ is the gain parameter, $a$ is the width of gain profile, $r = [(\varepsilon\eta)^2 + \zeta^2]^{1/2}$ with $\varepsilon$ being the parameter that defines the ellipticity of the gain profile, and $r_c$ sets the smaller semi-axis of the elliptical gain landscape (notice that the length of the longer semi-axis that is parallel to the $\eta$ axis depends on $r_c$ as well as on $\varepsilon$). Throughout the paper we set $r_c = 5.25$ and $a = 1.75$. For smaller values $r_c$ or $a$, higher gain levels are usually required for the generation of vortex solitons, otherwise all results reported remain qualitatively similar. In fact, one can always use scaling transformations in Eq. (1) to set a desired value for $r_c$. Also, we varied the parameter $\varepsilon$ in the range of $0.5 < \varepsilon < 1.0$.

Such 2D gain landscapes may be implemented experimentally by using properly shaped pump beams or materials with specially shaped concentration of active centers. Such fabrication techniques are well established and readily used in the area of fiber optics. Also, a novel method to manufacture laser gain media with a spatially variable gain profile, that may be used to implement the profiles we consider, has been invented [28].

Vortex solitons of Eq. (1) have the shape $q(\eta,\zeta,\xi)=[w_{\rm r}(\eta,\zeta)+iw_{\rm i}(\eta,\zeta)]\exp(ib\xi)$, with $b$ being the propagation constant, and $w_{\rm r}, w_{\rm i}$ being real and imaginary parts of the field. The topological charge of such vortex can be obtained as $m=(2\pi)^{-1}\oint \arctan(w_{\rm i}/w_{\rm r})d\phi$, where one can use any closed contour surrounding phase dislocation to calculate the accumulated phase $\phi$. In the case of $n$ dislocations the topological charges can be calculated for each of them. In Eq. (1) vortex solitons form due to the competition between gain and two-photon absorption that results in suppression of collapse and azimuthal instabilities even for cubic nonlinearities. In ring-like gain landscapes with $\varepsilon=1$ such solitons are attractors and form dynamically from various vortex-carrying inputs, such as $q|_{\xi=0}=Ar^{|m|}\exp(-r^2/w^2)\exp(im\phi)$, where $m$ is the input topological charge [24]. We study what happens with output vortex solitons upon stretching of the gain landscape (decrease of $\varepsilon$). Thus we solved Eq. (1) with the above mentioned inputs up to large distances $\xi \sim 10^4$. The stability and robustness of excitation of the resulting beams (attractors of the system) was tested by adding noise to the reached states and propagating them for another $\xi \sim 10^3$ units.

We found that while for $m=1$ inputs the phase dislocation always remains in the center of the vortex that becomes elliptical for $\varepsilon<1$, the input beams with $m>1$ always excite vortex solitons with $m$ spatially separated single-charge dislocations nested in the common vortex core. Figure 1 illustrates this effect for input beam with topological charge $m=2$ and $3$. Stretching of the gain landscapes results in a considerable expansion of the vortex states. High-charge dislocation located initially in the center of the pattern splits into $m$ dislocations and this splitting usually occurs in the direction of stretching of gain landscape. Notice that while each dislocation has unit charge, the total topological charge of the pattern calculated over the closed contour surrounding all dislocations remains equal to $m$. For vortices with odd charges one dislocation always remains in the center of the pattern, while in vortices with even charges all dislocations are shifted from the center.

The distance $d$ between outermost dislocations grows considerably with decrease of $\varepsilon$ (increase of ellipticity) and this spacing is always larger for vortices with higher $m$ values [Fig. 2(a)]. At the same time, the distance $d$ only slightly varies with modifications of gain and absorption - increasing gain results in the decrease of separation between dislocations. Still, the main factor determining the splitting of dislocations is the geometrical ellipticity of gain landscape. It should be stressed that the distances between neighboring dislocations are different for vortex solitons with topological charges $m>3$.

The dependence of energy flow $U = \int \int_{-\infty}^{\infty} |q|^2 \, d\eta d\zeta$ of vortex solitons on ellipticity parameter is shown in Fig. 2(b). The energy flow increases upon stretching of gain landscape. The curves $U(\varepsilon)$ are almost indistinguishable for $m=2,3$ and we show only one of them. The propagation constant for vortex soliton with two dislocations decreases with $\varepsilon$ [Fig. 2(c)].

While for solitons with $m=2,3$ presented above the phase dislocations were separated only along the $\eta$-axis (in the direction of stretching of gain landscape), it is also possible that dislocations shift along both $\eta$ and $\zeta$ axes. This is the case of high-charge vortex solitons with $m>3$. An example for $m=4$ is shown in Fig. 3. In this case the shift of dislocations may be so considerable that they fall into high-gain domains, around which the field intensity is not small anymore, that results in considerable deformation of vortex intensity distributions.

Similar effects may be also occur in fully three-dimensional gain landscapes localized in both space and time and moving with suitable group velocity, which may support stable dissipative vortex light bullets. Note that running gain propagating with the group velocity of the spatiotemporal wave packet has been achieved by tilted-pulse-front pumping approaches Ref. [29]. The method was introduced at the beginning of the 80s as a group-velocity matching technique between the optical pump pulse and the generated/amplified radiation. High gain and reduced amplified spontaneous emission may be obtained by using such traveling-wave excitation approach in laser materials, in particular dye solutions, semiconductors, and nonlinear crystals of the optical parametric amplifiers.

To study such possibility we considered the evolution of the wavepacket carrying phase dislocation in gain landscape with elliptical spatial and bell-like temporal shape that can be described by the three-dimensional version of Eq. (1):

$$i\frac{\partial q}{\partial \xi} = -\frac{1}{2}\left(\frac{\partial^2 q}{\partial \eta^2} + \frac{\partial^2 q}{\partial \zeta^2} + \frac{\partial^2 q}{\partial \tau^2}\right) - q|q|^2 + i\gamma(\eta,\zeta,\tau)q - i\alpha q|q|^2, \qquad (2)$$

where $\tau$ is the retarded time and we consider anomalous group-velocity dispersion regime. The gain landscape is described by the function $\gamma = p_i \exp[-(r-r_c)^2/a^2]\exp(-\tau^2/t^2)$, where $r = [(\varepsilon\eta)^2 + \zeta^2]^{1/2}$ and we selected $r_c = 3.75$, the width of gain landscape $a = 1.5$ and gain duration $t = 2$. We consider the situation when vortex-like structure can be observed in spatial domain, while temporal intensity distribution of the wavepacket remains bell-shaped. The topological charge of the pattern can be determined as before using a phase circulation over the closed contour in space surrounding phase dislocation.

Vortex light bullets were excited with input wavepackets described by the expression $q|_{\xi=0} = Ar^{|m|}\exp(-r^2-\tau^2)\exp(im\phi)$, where $\phi$ is the azimuthal angle in space, for different values of ellipticity parameter $\varepsilon$. As in two-dimensional configuration such inputs experience fast reshaping, emit radiation, and asymptotically approach stationary vortex light bullets in proper parameter

range. Here we consider only bullets with $m=2$ since solitons with higher topological charges require considerably higher gain levels for their stabilization.

Figure 4 confirms that the phenomenon studied in this paper occurs also in 3D geometries. The plot illustrates the gradual deformation of the output shape of vortex light bullet upon increase of the ellipticity of gain landscape. The progressive expansion of the vortex shape and separation of dislocations occurring in the direction of stretching of gain landscape are clearly visible. While in two-dimensional landscapes even strongly deformed vortex solitons remain stable, in three-dimensional landscapes the situation is different and strong stretching of gain profile results in destabilization of vortex bullet. Thus, for the parameters considered here stable vortex bullets with separated dislocations were found for $0.6<\varepsilon<1.0$. A typical dependence of separation between dislocations $d$ on ellipticity parameter is shown in Fig. 5(a). The separation $d$ monotonically increases width decrease of $\varepsilon$ and approaches zero when $\varepsilon \to 1$ for radially symmetric gain landscape. The total energy of vortex light bullet $U = \int\int\int_{-\infty}^{\infty} |q|^2 \, d\eta d\zeta d\tau$ as well as its propagation constant $b$ are the monotonically decreasing functions of the ellipticity parameter $\varepsilon$ [Fig. 5(b)].

Summarizing, we showed that instead of conventional high-charge vortex solitons with single phase dislocation in the center the elliptical gain landscapes can support stable two- and three-dimensional vortex states with multiple spatially separated dislocations. They are attractors of the system and can be excited with various vortex-carrying input beams. The separation between dislocations may be comparable with the width of amplifying ring and it is dictated mainly by the geometry of gain landscape. The phenomenon described here may be also relevant to all areas of physics where localized vortex-carrying states exist.

The work of O. V. Borovkova was supported by the Ministry of Science and Innovation, Government of Spain, grant FIS2009-09928.

**Figure captions**

Figure 1. (Color online) Field modulus (top) and phase (bottom) distributions for dissipative vortex solitons with two phase dislocations at (a) $\varepsilon=0.9$, (b) $\varepsilon=0.7$, (c) $\varepsilon=0.5$ and solitons with three phase dislocations at (d) $\varepsilon=0.9$, (e) $\varepsilon=0.7$, and (f) $\varepsilon=0.5$. In all cases $\alpha=2$ and $p_i=7$. All quantities are plotted in arbitrary dimensionless units.

Figure 2. (Color online) (a) The separation between outermost phase dislocations for vortex solitons with $n=2$ and $3$ versus $\varepsilon$. Energy flow (b) and propagation constant (c) of vortex soliton with $m=2$ versus $\varepsilon$. All quantities are plotted in arbitrary dimensionless units.

Figure 3. (Color online) Field modulus (top) and phase (bottom) distributions in unconventional azimuthally modulated dissipative vortex solitons with four phase dislocations at (a) $\varepsilon=0.9$, (b) $\varepsilon=0.7$, and (c) $\varepsilon=0.5$. In all cases $\alpha=2$ and $p_i=7$. All quantities are plotted in arbitrary dimensionless units.

Figure 4. Isosurface plots showing field modulus distribution in vortex light bullets with two phase dislocations for $\varepsilon=1$ (left), $\varepsilon=0.85$ (center), and $\varepsilon=0.7$ (right). In all cases $\alpha=3.5$, $p_i=9$.

Figure 5. (a) The separation between outermost phase dislocations for vortex light bullet with $m=2$ and (b) the energy of such bullets versus $\varepsilon$ at $\alpha=3.5$, $p_i=9$. All quantities are plotted in arbitrary dimensionless units.

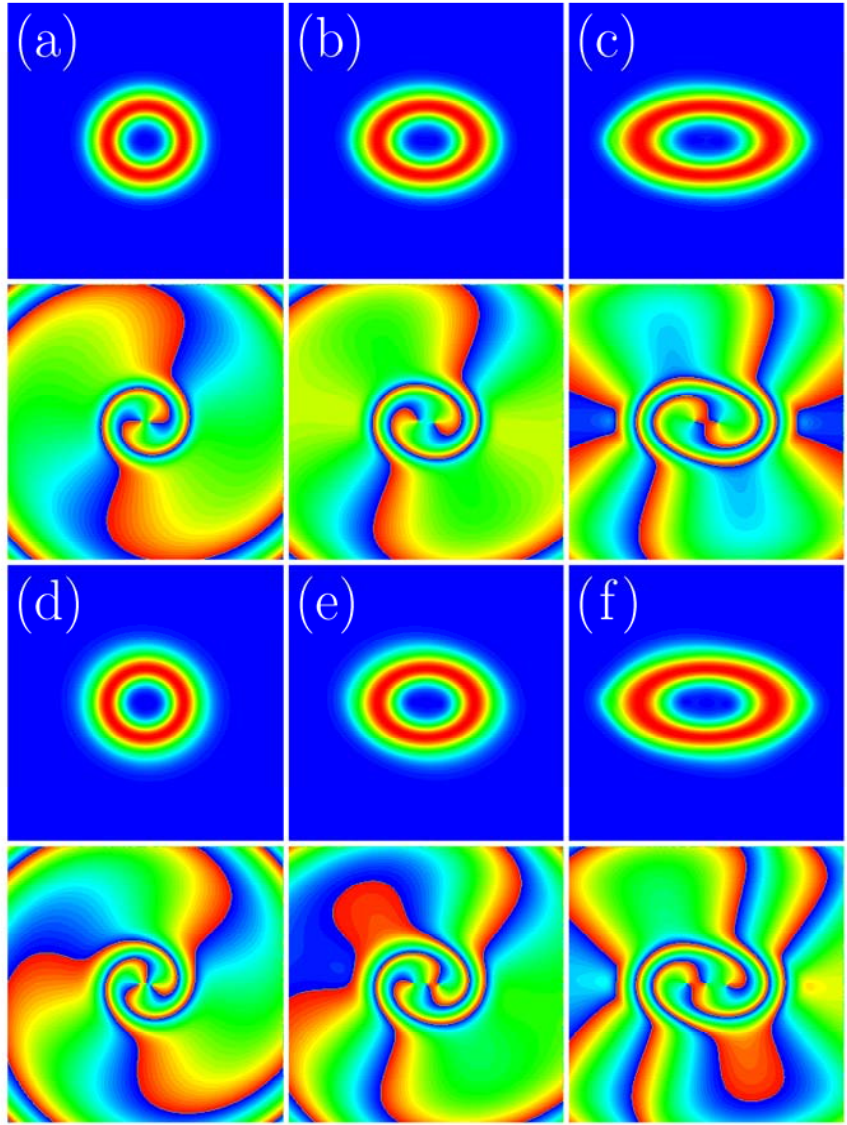

Figure 1. (Color online) Field modulus (top) and phase (bottom) distributions for dissipative vortex solitons with two phase singularities at (a) $\varepsilon = 0.9$, (b) $\varepsilon = 0.7$, (c) $\varepsilon = 0.5$ and solitons with three phase singularities at (d) $\varepsilon = 0.9$, (e) $\varepsilon = 0.7$, and (f) $\varepsilon = 0.5$. In all cases $\alpha = 2$ and $p_i = 7$. All quantities are plotted in arbitrary dimensionless units.

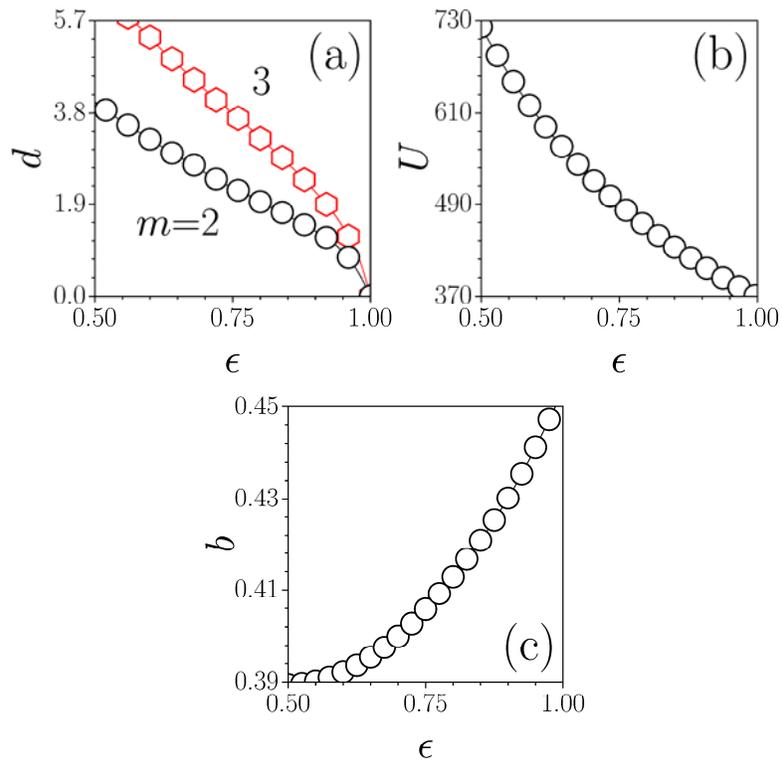

Figure 2. (Color online) (a) The separation between outermost phase singularities for vortex solitons with $m=2$ and 3 versus $\varepsilon$. Energy flow (b) and propagation constant (c) of vortex soliton with $m=2$ versus $\varepsilon$. All quantities are plotted in arbitrary dimensionless units.

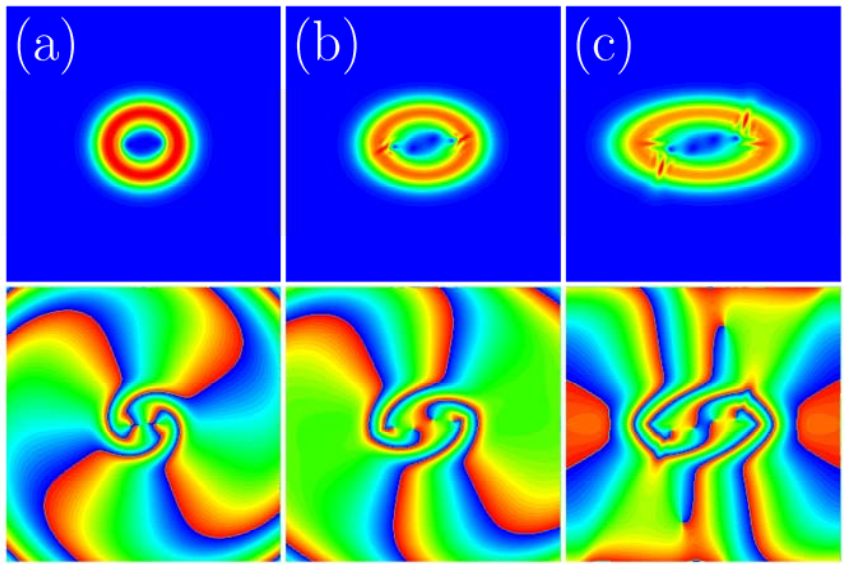

Figure 3. (Color online) Field modulus (top) and phase (bottom) distributions in unconventional azimuthally modulated dissipative vortex solitons with four phase singularities at (a) $\varepsilon = 0.9$, (b) $\varepsilon = 0.7$, and (c) $\varepsilon = 0.5$. In all cases $\alpha = 2$ and $p_\mathrm{i} = 7$. All quantities are plotted in arbitrary dimensionless units.

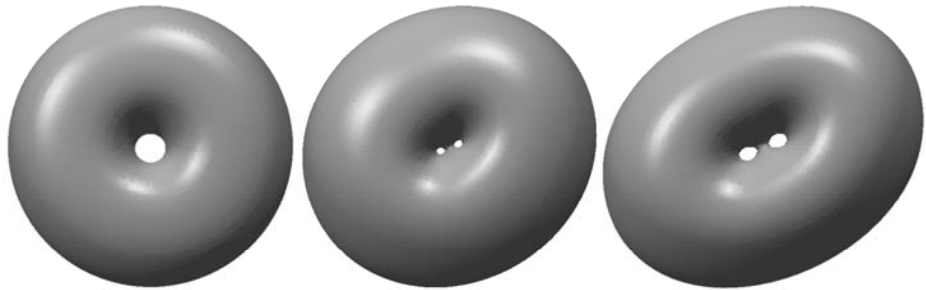

Figure 4. Isosurface plots showing field modulus distribution in vortex light bullets with two phase singularities for $\varepsilon=1$ (left), $\varepsilon=0.85$ (center), and $\varepsilon=0.7$ (right). In all cases $\alpha=3.5$, $p_{\mathrm{i}}=9$. All quantities are plotted in arbitrary dimensionless units.

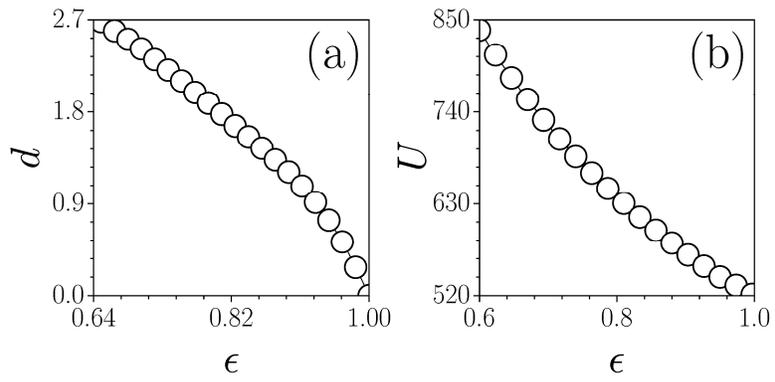

Figure 5. (a) The separation between outermost phase singularities for vortex light bullet with $m=2$ and (b) the energy of such bullets versus $\varepsilon$ at $\alpha=3.5$, $p_\mathrm{i}=9$. All quantities are plotted in arbitrary dimensionless units.